# Strong orientation dependent spin-orbit torque in antiferromagnet $Mn_2Au$

X. F. Zhou,[1,2] J. Zhang,[3] F. Li,[1,2] X. Z. Chen,[1,2] G. Y. Shi,[1] Y. Z. Tan,[1] Y. D. Gu,[1] M. S. Saleem,[1] H. Q. Wu,[2] F. Pan,[1,2] and C. Song[1,2,*]

[1] *Key Laboratory of Advanced Materials (MOE), School of Materials Science and Engineering, Tsinghua University, Beijing 100084, China*

[2] *Beijing Innovation Center for Future Chip, Tsinghua University, Beijing 100084, China.*

[3] *School of Physics and Wuhan National High Magnetic Field Center, Huazhong University of Science and Technology, Wuhan 430074, China*

Antiferromagnets with zero net magnetic moment, strong anti-interference and ultrafast switching speed have potential competitiveness in high-density information storage. Body centered tetragonal antiferromagnet $Mn_2Au$ with opposite spin sub-lattices is a unique metallic material for Néel-order spin-orbit torque (SOT) switching. Here we investigate the SOT switching in quasi-epitaxial (103), (101) and (204) $Mn_2Au$ films prepared by a simple magnetron sputtering method. We demonstrate current induced antiferromagnetic moment switching in all the prepared $Mn_2Au$ films by a short current pulse at room temperature, whereas different orientated films exhibit distinguished switching characters. A direction-independent reversible switching is attained in $Mn_2Au$ (103) films due to negligible magnetocrystalline anisotropy energy, while for $Mn_2Au$ (101) and (204) films, the switching is invertible with the current applied along the in-plane easy axis and its vertical axis, but becomes attenuated seriously during initially switching circles when the current is applied along hard axis, because of the existence of magnetocrystalline anisotropy energy. Besides the fundamental significance, the strong orientation dependent SOT switching, which was not realized irrespective of ferromagnet and antiferromagnet, provides versatility for spintronics.

---

* songcheng@mail.tsinghua.edu.cn

**I. INTRODUCTION**

Alternate arrangement of magnetic moments on adjacent atoms makes antiferromagnets (AFM) show zero net magnetization without stray field and resultant immunity to external perturbations [1]. After more than half century of passive role of AFM for the exchange bias effect at the interface coupled to ferromagnets (FM) [2], antiferromagnet spintronics have emerged as a fascinating research area and stimulated intense interest due to their potential for ultrafast and ultrahigh-density spintronics [1,3–7]. Much beyond the static behavior of AFM as supporting layer in spin valves and magnetic tunnel junctions [8], different methods have been demonstrated to manipulate the AFM spins in the AFM-based memory resistors [9] and tunneling anisotropic magnetoresistance [6,10], taking advantage of FM switching [11], field cooling [12], strain [13], electric field [14–18], and electric current which is realized recently [5,19,20].

Controlling magnetism by current, namely spin-transfer torque (STT) [21] and spin-orbit torque (SOT) [22–24], manifests great superiority for low-power spintronics, and favorable progress has been made in ferromagnetic (FM) systems and relevant magnetic random access memory (MRAM). The SOT has recently been used for the switching of both synthetic antiferromagnets [25,26] and antiferromagnetic CuMnAs [5,20] mainly attributed to the role of field-like torque $dM_{A,B}/dt \sim M_{A,B} \times p_{A,B}$, where the effective field proportional to $p_A = -p_B$ acting on the spin-sublattice magnetizations $M_{A,B}$ [4,5]. The SOT in collinear antiferromagnets have been studied in bulk $Mn_2Au$ tight-binding models. The $Mn_2Au$ crystal two sublattices connected by inversion and microscopic calculations based on the Kubo formula show that the Néel SOT in $Mn_2Au$ is predominantly of field-like character [4,24,27]. And current-driven SOT was also theoretically predicted in the easy-prepared metallic antiferromagnet $Mn_2Au$ with similar opposite spin sub-lattices with CuMnAs [1,4,24,27], which shows high Néel temperature above 1000 K [27], higher conductivity, and resultant lower energy dissipation for application, compared to the arsenic-based AFM, demonstrating pressing demand and great significance for the study of SOT in $Mn_2Au$.

The orientation dependent spin-orbit interaction and corresponding SOT have been commonly ignored mainly because the perpendicular magnetic anisotropy of the ferromagnetic stacks is persistently limited by a certain preferred orientation, such as (111)-oriented Pt/Co and (100)-oriented Ta/CoFeB. Recently, facet-dependent spin Hall angle was observed in triangular antiferromagnet $IrMn_3$, which provides efficient charge-to-spin current conversion and concomitant spin current propagating from the antiferromagnetic into an adjacent ferromagnetic material [28]. Considering that the field-like torque in AFM with opposite spin sub-lattices is not restricted by a certain crystalline orientation, the orientation dependent magnetic switching is expected in this scenario, which has not been realized both in ferromagnets and antiferromagnets, providing a versatile candidate for spintronics. Compared to polycrystalline $Mn_2Au$ films grown by molecular beam epitaxy [29,30] and partially (001)-oriented $Mn_2Au$ films deposited by sputtering [31], experiments below attain quasi-epitaxial (103), (101) and (204) $Mn_2Au$ films, and demonstrate the orientation dependent Néel-order spin-orbit torque switching in single layer $Mn_2Au$ films with different switching features for different orientations.

## II. METHODS

40 nm-thick (103)-, (101)-, (204)-oriented $Mn_2Au$ films, were grown on single crystal MgO (111), $SrTiO_3$ (100) and MgO (110) substrates, respectively, by magnetron sputtering at 300 ºC. The base pressure is $2 \times 10^{-5}$ Pa, and the growth rate is 0.12 nm/s using a $Mn_2Au$ alloy target (atomic ratio: 2:1). High resolution transmission electron microscopy (HRTEM) were carried out in JEM 2100. Magnetic properties were measured by a superconducting quantum interference device (SQUID) magnetometry. XAS/XMLD measurements in total electron yield mode were carried out at the Beamline BL08U1A in Shanghai Synchrotron Radiation Facility (at 300 K). The XAS spectra normalization was made by dividing the spectra by a factor such that the $L_3$ pre-edge and $L_2$ post-edge have identical intensities for the two polarizations. After that, the pre-edge spectral region was set to zero and the peak at the $L_3$ edge was set to one. Mn $L$-edge XMLD curves are characterized by the difference between linearly horizontal (//) and vertical (⊥) polarized x-ray absorption spectroscopy. The XMLD was measured at different in-plane rotation angles ($\theta$), where $\theta$ is the angle between a crystalline axis of Mn$_2$Au and horizontal polarization direction. Star devices for the SOT switching were fabricated using standard photolithography and argon ion milling procedures. The writing current pulses of ~$10^7$ A cm$^{-2}$ were generated by a Keysight B2961A source meter. After each writing pulse, a delay of 10 s was used for thermal relaxation, and then a measurement of the transversal voltage across the central part of the patterned structure was recorded when applying reading current with a density of ~$10^5$ A cm$^{-2}$ generated by Agilent B2901A source meter.

The magnetic anisotropy energy (MAE) for different orientated Mn$_2$Au was calculated by using standard Force Theory (FT). The lattice constants of Mn$_2$Au on different substrates used in the calculations were obtained by XRD experiments. The atomic position were fully relaxed. First a self-consistence of electron charge density (electronic potential) was performed in the absence of spin-orbit coupling, followed by a one-step calculation with the presence of spin-orbit coupling when the magnetization was aligned along different axes. The band energy difference was evaluated as the MAE.

## III. RESULTS

We first show the microstructure characterizations of 40 nm-thick Mn$_2$Au films in Fig. 1 with different epitaxial orientation grown by magnetron sputtering. X-ray diffraction (XRD) spectra show the (103), (101) and (204) textures for the Mn$_2$Au films grown on MgO (111), SrTiO$_3$ (100) and MgO (110) substrates, respectively, in left panel of Figs. 1(a)–1(c). All the films are single phase with no evidence of diffraction peak from second phases or other crystalline faces within the sensitivity of XRD measurements. Corresponding high resolution transmission electron microscopy (HRTEM) images of the $Mn_2Au$/substrate cross-sections are depicted in right panel of Fig. 1. There is no observation for grain boundaries in the overall of films, though some point defects still exist in the crystalline lattice, indicating the quasi-epitaxial growth mode for the present $Mn_2Au$ films by a simple sputtering method. The epitaxial growth mode provides a pre-condition for the observation of orientation dependent spin-orbit torque of $Mn_2Au$ films as discussed below.

We use a superconducting quantum interference device (SQUID) magnetometry to check antiferromagnetism of the $Mn_2Au$ films. Magnetic field was applied in plane, i.e. parallel to MgO substrate (111) plane, [110] edge. Figure 2(a) presents the magnetization curves of the $Mn_2Au$ (103) films at 300 K. Typical diamagnetic signals up to 50 kOe are observed, which reflects the diamagnetic feature of the substrate, indicating the antiferromagnetism of the $Mn_2Au$ (103) films without any ferromagnetic signal. $Mn_2Au$ (101) and (204) films show similar behaviors (see Figs. S1 and S2 of Ref. [32] for magnetization measurements). Moreover, 5 nm-thick permalloy is deposited on top of the 10 nm $Mn_2Au$ to reaffirm the antiferromagnetism of $Mn_2Au$ (103) films. Corresponding magnetization data of $Mn_2Au$ (10 nm) / permalloy (5 nm) at 20 K, after 10 kOe field-cooling from 300 K, are depicted in Fig. 2(b). The hysteresis loop exhibits an apparent shift of 150 Oe compared to its coercivity of 400 Oe, indicating the antiferromagnetic order of the $Mn_2Au$ (103) films. We then display a schematic of 3 × 3 unit cells of $Mn_2Au$ ($Mn_2Si$-type tetragonal structure with $a = b = 0.333$ nm and $c = 0.854$ nm) [33] in Fig. 2(c), where the $Mn_2Au$ (103) plane is highlighted with the horizontal direction of [100] (*a*-axis) and its vertical direction [0$\bar{3}$1]. Also visible is the neighboring sublattice with opposite magnetic moments, which is the prerequisite for the current driven magnetic switching in $Mn_2Au$ through spin-orbit torque.

To detect the switching of $Mn_2Au$ magnetic axis, the planar Hall resistance is obtained perpendicular to the reading current after two orthogonal writing current injecting to the device alternatively. Figure 3(a) depicts a schematic of the star device and current switching measurement. For this experiment, five successive 1 ms-width writing current pulses of $J_{write}$ = 2.10 × 10$^7$ A cm$^{-2}$ are applied along [010] axis (red arrow, referred to as $J_{write}$1) and then along its vertical direction, [$\bar{3}$01] axis (blue arrow, $J_{write}$2) [Figs. 3(a) and 3(b)]. The current density is calculated with the following parameters: 42 mA applied writing current, 40 nm-thick Mn$_2$Au films and 5 μm-width of the writing current arm of the star device. Ten seconds are recorded after each writing current pulse to release possible tiny heat effect caused by the short current pulse of 1 ms. Then a reading current of $J_{read}$ = 4.17 × 10$^5$ A cm$^{-2}$ is applied 45° away from the writing current, while the planer Hall resistance is recorded.

The [010]-directed writing pulses are expected to align a preference of domains with AFM spin axis perpendicular to [010]-axis, then the writing current along [$\bar{3}$01] direction drives the magnetic switching back to [010]-axis. Following this concept, Figure 3(c) shows the Hall resistance ($R_{Hall}$) variation as a function of the number of current pulses, five successive current pulses along [010] and [$\bar{3}$01] axes alternatively [Fig. 3(b)]. The five current pulses along [010] axis ($J_{write}$1) set the AFM spins axis perpendicular to [010] axis, which gradually increases $R_{Hall}$ approximately 2.8 mΩ (red squared line). Remarkably, a current pulse as short as 1 ms along [$\bar{3}$01] ($J_{write}$2) drives the AFM spins rotating back to [010] direction, resulting in the $R_{Hall}$ decreased 4.4 mΩ to about −1.6 mΩ in the opposite sign. The following four pulses make the variation of $R_{Hall}$ tend to be saturated (blue squared line). After that, the five current pulses along [010] axis increase $R_{Hall}$, which make $R_{Hall}$ reverse back to 2.6 mΩ, in the vicinity of initial 2.8 mΩ. Note that a constant background is subtracted. Such a switching experiment is conducted 10 circles. Apparently, $R_{Hall}$ increase and decrease without clear deterioration, indicating the present magnetic switching is reversible, repeatable, and reliable.

The abrupt change of Hall resistance with the first current pulse, followed by gradually enhanced Hall resistance in the following four current pulses, reflects the multi-domain switching feature of the present Néel-order SOT, which is quite characteristic for SOT in both ferromagnetic system [34] and AFM CuMnAs [5]. A temperature calibration was used to monitor the stability of the test temperature at 300 K, excluding the variation of resistance caused by temperature. It is worth to mention that the Hall resistance is independent of the direction of writing current, coinciding with the theory of current-driven switching of AFM [4]. Interestingly, the present switching is completed by a short current pulse of 1 ms, compared to the 50 ms pulse used in CuMnAs system [5], which reflects strong switching capability of metallic AFM system.

We then investigate the writing current density and current duration time dependent switching of the $Mn_2Au$ films. Corresponding data are displayed in Figs. 3(d) and 3(e), respectively. For these experiments, different writing current densities ($2.25 \times 10^7$, $2.40 \times 10^7$, $2.55 \times 10^7$ A $cm^{-2}$) and current pulses width (3, 5, and 15 ms) are applied to the star device. The measurement sequence is identical to that of Fig. 3(c). With increasing writing current density and current duration time, the magnitude of $R_{Hall}$ becomes larger, most likely due to stronger SOT switching by the applied current and the resultant 90° switching of AFM moment. According to the current induced switching results, the first and second pulses are decisive which result in sharp Hall resistance change and major magnetic moments switching, while following writing pulses causes relatively small magnetic moments switching and Hall resistance variety. And the switching is reproducible during the five cycles. It should be mentioned that a much higher writing current and pulses width would induce non-recyclable switching or device rupture by the thermal or electromigration effect. The current induced switching results of different thicknesses point out no thickness-dependent switching behavior, reflecting the calculated inherent switching ability of bulk $Mn_2Au$ (see Fig. S3 of Ref. [32] for thickness-dependent switching of $Mn_2Au$) [4,24,27].

In general, the antiferromagnets show magnetic anisotropy, which might affect the SOT switching. Thus the switching capability of the $Mn_2Au$ (103) films should be dependent on the current applied in different crystalline axes, besides two typical axes, [010] and [$\bar{3}$01]. We then explore the SOT effect when both of the writing current and reading current directions are rotated 45° from their original counterparts. In this case, the direction of the writing current here is along the direction of reading current in Fig. 3(a). It is interesting to find in Fig. 3(f) that $R_{Hall}$ also apparently increase and decrease without clear deterioration, especially the recycling behavior for this switching is reliable. Also, the magnitude of $R_{Hall}$ for the SOT switching is comparable to its counterpart in Fig. 3(c). The comparable SOT switching behaviors in Figs. 3(a) and 3(c) suggest that the $Mn_2Au$ (103) films have no apparent antiferromagnetic easy axis.

The SOT switching experiments have revealed that the AFM spins tend to arrange randomly and are comparable in each direction of (103) orientation $Mn_2Au$. We now address the question whether an analogous spin structure can also be detected by x-ray magnetic linear dichroism (XMLD), which stands out as a unique method to identify the spin orientations in AFM despite the absence of net moment [35,36]. Figure 4 presents Mn *L*-edge XMLD curves, characterized by the difference between linearly horizontal (//) and vertical ($\perp$) polarized x-ray absorption spectroscopy (XAS; see Fig. S4 of Ref. [32] for the original XAS). The intensity of XMLD is determined by the difference between angles of AFM spin/horizontal polarization and AFM spin/vertical polarization: the larger the angle difference is, the stronger the XMLD signal is.

To understand the alignment of AFM spins, the XMLD at several typical angles ($\theta$) are measured by changing in-plane angle $\theta$: the angle between [010] axis of $Mn_2Au$ and horizontal polarization direction, as displayed in the inset of Fig. 4. For $\theta = 0°$, the horizontal polarized x-ray is set to [010] axis of the $Mn_2Au$ (103) films, while the vertical polarized x-ray is along its vertical direction. The obtained XMLD signal is rather weak without characteristic "positive to negative" XMLD pattern at $L_3$-edge (about 635–645 eV) [35]. Similar XMLD signals are observed for the spectra measured with $\theta = 30°$, 45°, and 60°, three typical directions between [010] and [$\bar{3}$01] axes (see Fig. S5 of Ref. [32] for the pole figure of Mn$_2$Au (103) films). This finding demonstrates the absence of dominant orientation of the ordered spin textures (Néel vectors) at Mn$_2$Au (103) plane. This feature guarantees that the magnetic switching of Mn$_2$Au (103) is mainly determined by the vector of field-like torque, which realigns the spin texture perpendicular to the current channel on the basis of the theory [4]. The current applied alternatively at [010] and [$\bar{3}$01] axes lead to the reversible modulation of AFM spins and the resultant planar Hall effect.

We then turn towards the SOT switching of Mn$_2$Au (101) films. The star devices for the Mn$_2$Au (101) films and measurement setup are identical to those of Mn$_2$Au (103) in Fig. 3(a), but the crystal axis at $\theta = 0°$ is [13$\bar{1}$] and its vertical axis is [$\bar{1}$31] (see Fig. S6 of Ref. [32] for the pole figure of Mn$_2$Au (101) films). Figure 5(a) shows the Hall resistance variation when four successive writing current pulses are applied alternatively along [13$\bar{1}$] (red squares) and [$\bar{1}$31] (blue squares) of the Mn$_2$Au (101) films. Note that $R_{\text{Hall}}$ changes with the direction of 1 ms-width writing current pulse ($J_{\text{write}} = 2.00 \times 10^7$ A cm$^{-2}$) for the first circle. However, this variation attenuates seriously after the first circle, and almost vanishes after five circles. These features reflect that the current can not reversibly switch the AFM spin texture between [13$\bar{1}$] and [$\bar{1}$31] axes. The situation turns out to be dramatically different when the probe configuration rotates 45°. That is, both of the writing current and reading current are rotated 45° from its origin case in Fig. 5(a). Corresponding data are presented in Fig. 5(b). $R_{\text{Hall}}$ can be reversibly modulated by the writing current pulse of 1 ms-width and $J_{\text{write}} = 2.00 \times 10^7$ A cm$^{-2}$. This characterization indicates that the AFM spin texture is switched between [010] axis ($\theta = 45°$) and its vertical direction.

The transport measurements suggest that the spin texture of the Mn$_2$Au (101) films should be different from that of the Mn$_2$Au (103) films. Figure 5(c) depicts the XMLD spectra of three typical angles $\theta = 0°$, 30°, and 45°, where the angle strongly affects the XMLD signal.

For $\theta = 0°$, the horizontal and the vertical polarized x-ray is set to $[13\bar{1}]$ and $[\bar{1}31]$ axes of the Mn$_2$Au (101) films, respectively. It is found that the XMLD signal at Mn $L_3$-edge is quite weak, indicating no obvious difference between $[13\bar{1}]$ and $[\bar{1}31]$. An analogous XMLD spectrum is obtained for the $\theta = 30°$ but the magnitude of XMLD signals is somehow enhanced. The case differs completely as $\theta = 45°$. A characteristic positive to negative XMLD pattern at $L_3$-edge indicates that the distinguished spin textures between these two orientations, [010] axis ($\theta = 45°$) and $[\bar{1}01]$ axis ($\theta = 135°$). Because of the broken symmetry in the strained Mn$_2$Au films, the Néel vector orientation is most likely aligned along [010] ($\theta = 45°$) [7], making [010] an easy axis for the as-grown Mn$_2$Au (101) films. With the emergence of SOT by the writing current pulse along [010] direction, the AFM spin texture can be switched to its vertical direction, and then switch reversibly and flexibly between these two directions by applying alternative current pulses. However, for the hard axis of $[13\bar{1}]$ ($\theta = 0°$) or $[\bar{1}31]$ ($\theta = 90°$), the Néel vector orientation is difficult to be arranged in these energy disfavored directions. Even a part of spin texture is reluctant to be switched to these two directions by SOT, whereas the switching is hard to maintain, because the AFM spins are pinned gradually by the easy axis [010] at $\theta = 45°$.

We then show the SOT switching and spin texture measurements of Mn$_2$Au (204) films in Fig. 6. The experiments are carried out using the same procedure as that of Mn$_2$Au (103) films in Fig. 3. One can see in Fig. 6(a) that the variation of Hall resistance is apparent when applying four successive 1 ms-width writing current pulses of $J_{\text{write}} = 2.00 \times 10^7$ A cm$^{-2}$ along [010] ($\theta = 0°$, red squares) and $[\bar{2}01]$ axes ($\theta = 90°$, blue squares) alternatively (see Fig. S7 of Ref. [32] for the pole figure of Mn$_2$Au (204) films). Applied current along [010] initially aligned the AFM moments perpendicular to current direction by the SOT effect and $R_{\text{Hall}}$ turns out to increase. Then $R_{\text{Hall}}$ turns out to decrease with the current along $[\bar{2}01]$, corresponding to the AFM switching back to [010] axis. Two successive current pulses would saturate the variation of $R_{Hall}$ to some extent. Remarkably, $R_{Hall}$ increase and decrease with ten series of current pulses along [010] and [$\bar{2}$01], demonstrating the AFM moments are switched reversibly between two perpendicular directions in ten circles. The switching feature is completely different when the measurement configuration rotates 45°, i.e., the writing current pulses are applied at $\theta$ = 45°. The variation of $R_{Hall}$ is greatly reduced with increasing the cycling and keeps unchanged after three switching circles. This feature reveals that the magnetic moments could not be switched effectively in these two directions at $\theta$ = 45° and 135°.

XMLD signals in Fig. 6(c) support the transport measurements. For $\theta$ = 0°, the horizontal and the vertical polarized x-ray is set to [010] and [$\bar{2}$01] axes of the Mn$_2$Au (204) films, respectively. Interestingly, the XMLD signal at $\theta$ = 0° features as a characteristic positive to negative XMLD pattern at $L_3$-edge, reflecting the easy axis along [010] for the as-grown Mn$_2$Au (204) films with strain induced by the MgO (110) substrates [37]. Accordingly, the XMLD signals at $\theta$ = 30° are greatly reduced, followed by the absence of XMLD $L_3$-edge signal at $\theta$ = 45°, reflecting that the spin textures strongly prefer to be aligned along the easy axis [010]. The Néel vectors are switched between [010] and its vertical direction. Such a strong AFM anisotropy is in a good agreement with the efficient switching between these two directions, followed by the comparatively easy SOT switching and speedy saturation of $R_{Hall}$ in Fig. 6(a). In contrast, the SOT switching in non-easy axis becomes more difficult with fast attenuated variation of $R_{Hall}$ in Fig. 6(b).

In addition, the Joule heat produced by a pulse current electric current is simulated using finite element modelling by COMSOL software. The instantaneous maximum temperature of Mn$_2$Au device can increase from 300 K to 424 K after a pulse of $2 \times 10^7$ A cm$^{-2}$ amplitude and 1 ms length, which is far below the Néel temperature of Mn$_2$Au (>1000 K). Nevertheless, heat assisted switching could play a role on reducing the current density, from theoretical ~$10^8$–$10^9$ A cm$^{-2}$ [4] to ~$10^7$ A cm$^{-2}$. The electric current induced Oster field distribution is also simulated using finite element modelling by COMSOL software. The maximum Oster field is only ~100 Oe. Such a disturbance of the magnetic field can be ignored for antiferromagnetic $Mn_2Au$ (see Figs. S8 and S9 of Ref. [32] for the Joule heat and Oster field simulation).

## IV. DISCUSSION

In order to understand the switching behaviour in different orientated $Mn_2Au$, the magnetic anisotropy energy (MAE) of different orientated $Mn_2Au$ is calculated by using standard Force Theory (FT). According to the symmetry of bulk $Mn_2Au$, the MAE can be expressed by [7],

$$E(\omega, \varphi) = K_{2\perp} \sin^2 \omega + K_{4\perp} \sin^4 \omega + K_{4\parallel} \sin^4 \omega \cos 4\varphi \qquad (1)$$

here $\omega$ and $\varphi$ are the angles between magnetization ($M$) and [001], as well as the projection direction of $M$ on $Mn_2Au$ (001) plane and [100] axis, respectively; $K_{2\perp}$, $K_{4\perp}$, and $K_{4//}$ parameterize the uniaxial MAE constant, the fourth-order out-of-plane and in-plane MAE constants, respectively. Based on Eq. (1), the specific axes for the MAE are listed as follows: $E([001]) = E(0, \varphi) = 0$, $E([100]) = E(\pi/2, 0) = K_{2\perp} + K_{4\perp} + K_{4//} = E([010]) = E(\pi/2, \pi/2)$, and $E([110]) = E(\pi/2, \pi/4) = K_{2\perp} + K_{4\perp} - K_{4//} = E([\bar{1}10]) = E(\pi/2, 3\pi/4)$. Apparently, the easy axis is determined by the sign of $K_{4//}$. The easy axis is along [110] when $K_{4//} > 0$ and [100] when $K_{4//} < 0$. The lattice constants of different orientated-$Mn_2Au$ used in the calculations are obtained by XRD experiments. Through first-principles calculations, the MAE constants $K_{2\perp}$, $K_{4\perp}$, and $K_{4//}$ can be obtained according to the lattice constant and orientation (see Table S1 of Ref. [32] for the MAE constants), giving rise to the angular dependence of MAE in Fig. 7(a).

Note that MAE decreases with increasing angle $\omega$, and becomes the lowest level for $\omega = 90°$, corresponding to $Mn_2Au$ (001) plane, which is the easy magnetized plane. Concerning the details in (001) easy plane, (103) orientated $Mn_2Au$ exhibits negligible difference of MAE, accompanied by comparable MAE value at $\varphi = 45°$ ([110]) and $\varphi = 90°$ ([010] direction), thus the magnetic moments of $Mn_2Au$ (103) can be switched by current either between [100] and [010] or between [110] and [$\bar{1}$10], as illustrated in Fig. 3. The situation differs abruptly for (101) and (204) orientated $Mn_2Au$, MAE is different when the magnetization rotates within the (001) plane, associated with lower MAE at $\varphi = 0°$ ([100]) and 90° ([010]) than that at $\varphi = 45°$ ([110]) and 135° ([$\bar{1}$10]). Consequently, the magnetic moments of (101) and (204) orientated $Mn_2Au$ can be switched by current between [100] and [010] rather than [110] and [$\bar{1}$10], as displayed in Figs. 5 and 6. An inspection of the MAE of $Mn_2Au$ (103), (101) and (204) planes in Fig. 7(b) shows that [010] direction ($\theta = 0°$) is indeed the energy minimum, whereas its vertical directions, [$\bar{3}$01], [$\bar{1}$01] and [$\bar{2}$01] for (103), (101) and (204) planes are energy maximum, respectively. In this senario, it is difficult for the magnetic moments switches between [010] and these directions. Instead, the magnetic moments are switched by two orthogonal currents between two easy axes, one is [010] lying within (103), (101) and (204) planes, while the other easy axis [100] lying out of these planes, as presented in Fig. 7(c). In fact, [$\bar{3}$01], [$\bar{1}$01] and [$\bar{2}$01] can be considered at the respective projection of [100] direction in (103), (101) and (204) planes of $Mn_2Au$, which are parallel to their substrate surface. The variation of Hall resistances are read out by the planar Hall effect in (103), (101) and (204) planes, respectively.

## V. CONCLUSION

We have demonstrated the spin-orbit torque in body centered tetragonal antiferromagnet $Mn_2Au$ with opposite spin sub-lattices. There are five main features for the present findings: (i)

The SOT switching behavior shows strong orientation dependence. The AFM moments are switched reversibly when the crystalline orientation plane has no strong magnetic anisotropy or the writing current is applied along the easy axis [010] and its vertical direction (the (101) and (204) orientation in-plane projection of another easy axis [100]), which is not only a typical feature for the SOT in AFM from the fundamental viewpoint, but also provides a versatile candidate for antiferromagnet spintronics; (ii), $Mn_2Au$ was theoretically to realize sizable reorientations at current densities of $\sim 10^8$–$10^9$ A cm$^{-2}$ [4]. We demonstrate the switching of $Mn_2Au$ magnetic moments at current densities of $\sim 10^7$ A cm$^{-2}$, ascribed to multi-domain wall and heat assisted switching; (iii) The growth parameter of quasi-epitaxial $Mn_2Au$ films by an easy access magnetron sputtering method has been optimized, where the buffer layer and molecular beam epitaxy are not necessary; (iv) The SOT switching has been realized in CuMnAs (arsenic-based semimetal), but only magnetic metals and alloys are generally considered as the electrode of magnetic tunnel junctions and relevant spintronics. Our work might advance the use of AFM alloys as a functional layer in spintronics; (v) The SOT switching of antiferromagnet occurs at room temperature, which is critical for the application of AFM spintronics. Our findings not only add a new dimension to spin-orbit torque, but also represent a promising step towards antiferromagnet spintronics.

**ACKNOWLEDGEMENTS**

We are grateful to Dr. Luqiao Liu of MIT for fruitful discussions and criticial reading of the manuscript. We acknowledge the Beamline BL08U1A in Shanghai Synchrotron Radiation Facility (SSRF) for XAS/XMLD measurements and Center for Testing and Analyzing of Materials for technical support. C.S. acknowledges the support of Young Chang Jiang Scholars Program. This work was supported by the National Natural Science Foundation of China (Grant Nos. 51671110, 51571128) and the National Key R&D Program of China (Grant Nos. 2017YFB0405704).

**Note added.**

During the preparation of this manuscript, we found two relevant works reporting current induced switching in Mn$_2$Au [38,39]. They reported the SOT switching of Mn$_2$Au (001).

[1]  T. Jungwirth, X. Marti, P. Wadley, and J. Wunderlich, Antiferromagnetic spintronics, Nat. Nanotechnol. **11**, 231-241 (2016).

[2]  W. Zhang and K. M. Krishnan, Epitaxial exchange-bias systems: From fundamentals to future spin-orbitronics, Mater. Sci. Eng. R **105**, 1-20 (2016).

[3]  A. H. MacDonald and M. Tsoi, Antiferromagnetic metal spintronics, Phil. Trans. R. Soc. A **369**, 3098–3114 (2011).

[4]  J. Železný, H. Gao, K. Výborný, J. Zemen, J. Mašek, Aurélien Manchon, J. Wunderlich, Jairo Sinova, and T. Jungwirth, Relativistic Néel-order felds induced by electrical current in antiferromagnets, Phys. Rev. Lett. **113**, 157201 (2014).

[5]  P. Wadley, B. Howells, J. Železný, C. Andrews, V. Hills, R. P. Campion, V. Novák, K. Olejník, F. Maccherozzi, S. S. Dhesi, S. Y. Martin, T. Wagner, J. Wunderlich, F. Freimuth, Y. Mokrousov, J. Kuneš, J. S. Chauhan, M. J. Grzybowski, A. W. Rushforth, K. W. Edmonds, B. L. Gallagher, and T. Jungwirth, Electrical switching of an antiferromagnet, Science **351**, 587–590 (2016).

[6]  Y. Y. Wang, C. Song, B. Cui, G. Y. Wang, F. Zeng, and F. Pan, Room-temperature perpendicular exchange coupling and tunneling anisotropic magnetoresistance in antiferromagnet-based tunnel junction, Phys. Rev. Lett. **109**, 137201 (2012).

[7]  A. B. Shick, S. Khmelevskyi, O. N. Mryasov, J. Wunderlich, and T. Jungwirth, Spin-orbit coupling induced anisotropy effects in bimetallic antiferromagnets: A route towards antiferromagnetic spintronics, Phys. Rev. B **81**, 212409 (2010).

[8]  S. Ikeda, K. Miura, H. Yamamoto, K. Mizunuma, H. D. Gan, M. Endo, S. Kanai, J. Hayakawa, F. Matsukura, and H. Ohno, A perpendicular-anisotropy CoFeB–MgO

magnetic tunnel junction, Nat. Mater. **9**, 721-724 (2010).

[9] X. Marti, I. Fina, C. Frontera, J. Liu, P. Wadley, Q. He, R. J. Paull, J. D. Clarkson, J. Kudrnovský, I. Turek, J. Kuneš, D. Yi, J. H. Chu, C. T. Nelson, L. You, E. Arenholz, S. Salahuddin, J. Fontcuberta, T. Jungwirth, and R. Ramesh, Room-temperature antiferromagnetic memory resistor, Nat. Mater. **13**, 367-374 (2014).

[10] B. G. Park, J. Wunderlich, X. Martí, V. Holý, Y. Kurosaki, M. Yamada, H. Yamamoto, A. Nishide, J. Hayakawa, H. Takahashi, A. B. Shick, T. Jungwirth, A spin-valve-like magnetoresistance of an antiferromagnet-based tunnel junction. Nat. Mater. **10**, 347-351 (2011).

[11] A. Scholl, M. Liberati, E. Arenholz, H. Ohldag, and J. Stöhr, Creation of an antiferromagnetic exchange spring, Phys. Rev. Lett. **92**, 247201 (2004).

[12] X. Z. Chen, J. F. Feng, Z. C. Wang, J. Zhang, X. Y. Zhong, C. Song, L. Jin, B. Zhang, F. Li, M. Jiang, Y. Z. Tan, X. J. Zhou, G. Y. Shi, X. F. Zhou, X. D. Han, S. C. Mao, Y. H. Chen, X. F. Han, and F. Pan, Tunneling anisotropic magnetoresistance driven by magnetic phase transition, Nat. Commun. **8**, 449 (2017).

[13] R. O. Cherifi, V. Ivanovskaya, L. C. Phillips, A. Zobelli, I. C. Infante, E. Jacquet, V. Garcia, S. Fusil, P. R. Briddon, N. Guiblin, A. Mougin, A. A. Ünal, F. Kronast, S. Valencia, B. Dkhil, A. Barthélémy, and M. Bibes, Electric-field control of magnetic order above room temperature, Nat. Mater. **13**, 345-351 (2014).

[14] J. Nogués, and I. K. Schuller, Exchange bias, J. Magn. Magn. Mater. **192**, 203–232 (1999).

[15] A. Scholl, M. Liberati, E. Arenholz, H. Ohldag, and J. Stöhr, Creation of an antiferromagnetic exchange spring, Phys. Rev. Lett. **92**, 247201 (2004).

[16] Y. Y. Wang, X. Zhou, C. Song, Y. N. Yan, S. M. Zhou G. Y. Wang, C. Chen, F. Zeng, and F. Pan, Electrical control of exchange spring in antiferromagnetic metals, Adv. Mater. **27**, 3196–3201 (2015).

[17] T. Zhao, A. Scholl, F. Zavaliche, K. Lee, M. Barry, A. Doran, M. P. Cruz, Y. H. Chu, C.

Ederer, N. A. Spaldin, R. R. Das, D. M. Kim, S. H. Baek, C. B. Eom, and R. Ramesh, Electrical control of antiferromagnetic domains in multiferroic $BiFeO_3$ films at room temperature, Nat. Mater. **5**, 823-829 (2006).

[18] C. Song, B. Cui, F. Li, X. Zhou, and F. Pan, Recent progress in voltage control of magnetism: Materials, mechanisms, and performance, Prog. Mater. Sci. **87**, 33–82 (2017).

[19] C. Song, Y. F. You, X. Z. Chen, X. F. Zhou, Y. Y. Wang and F. Pan, How to manipulate magnetic states of antiferromagnets, Nanotechnology **29**, 112001 (2018).

[20] M. J. Grzybowski, P. Wadley, K. W. Edmonds, R. Beardsley, V. Hills, R. P. Campion, B. L. Gallagher, J. S. Chauhan, V. Novak, T. Jungwirth, F. Maccherozzi, and S. S. Dhesi, Imaging Current-Induced Switching of Antiferromagnetic Domains in CuMnAs, Phys. Rev. Lett. **118**, 57701 (2017).

[21] D. C. Ralpha, and M. D. Stilesb, Spin transfer torques, J. Magn. Magn. Mater. **320**, 1190-1216 (2008).

[22] L. Liu, C. F. Pai, Y. Li, H. W. Tseng, D. C. Ralph, and R. A. Buhrman, Spin-torque switching with the giant spin Hall effect of tantalum, Science **336**, 555-558 (2012).

[23] Y. B. Fan, P. Upadhyaya, X. F. Kou, M. R. Lang, S. Takei, Z. Wang, J. S. Tang, L. He, L. T. Chang, M. Montazeri, and K. L. Wang. Magnetization switching through giant spin-orbit torque in a magnetically doped topological insulator heterostructure, Nature Mater. **13**, 699-704 (2014).

[24] J. Železný, H. Gao, A. Manchon, F. Freimuth, Y. Mokrousov, J. Zemen, J. Mašek, J. Sinova, and T. Jungwirth, Spin-orbit torques in locally and globally noncentrosymmetric crystals: Antiferromagnets and ferromagnets, Phys. Rev. B **95**, 014403 (2017).

[25] C. Bi, H. Almasi, K. Price, T. Newhouse-Illige, M. Xu, S. R. Allen, X. Fan, and W. Wang, Anomalous spin-orbit torque switching in synthetic antiferromagnets, Phys. Rev. B **95**, 104434 (2017).

[26] G. Y. Shi, C. H. Wan, Y. S. Chang, F. Li, X. J. Zhou, P. X. Zhang, J. W. Cai, X. F. Han, F.

Pan, and C. Song, Spin-orbit torque in MgO/CoFeB/Ta/CoFeB/MgO symmetric structure with interlayer antiferromagnetic coupling, Phys. Rev. B **95**, 104435 (2017).

[27] V. M. T. S. Barthem, C. V. Colin, H. Mayaffre, M. H. Julien, and D. Givord, Revealing the properties of $Mn_2Au$ for antiferromagnetic spintronics, Nat. Commun. **4**, 2892 (2013).

[28] W. Zhang, W. Han, S. H. Yang, Y. Sun, Y. Zhang, B. Yan, and S. S. P. Parkin, Giant facet dependent spin-orbit torque and spin Hall conductivity in the triangular antiferromagnet $IrMn_3$, Sci. Adv. **2**, e1600759 (2016).

[29] H. C. Wu, Z. M. Liao, R. G. Sumesh Sofin, G. Feng, X. M. Ma, A. B. Shick, O. N. Mryasov, and I. V. Shvets, $Mn_2Au$: Body-Centered-Tetragonal Bimetallic Antiferromagnets Grown by Molecular Beam Epitaxy, Adv. Mat. **24**, 6374 (2012).

[30] H. C. Wu, M. Abid, A. Kalitsov, P. Zarzhitsky, M. Abid, Z. M. Liao, C. Ó Coileáin, H. Xu, J. J. Wang, H. Liu, O. N. Mryasov, C. R. Chang, and I. V. Shvets, Anomalous Anisotropic Magnetoresistance of Antiferromagnetic Epitaxial Bimetallic Films: $Mn_2Au$ and $Mn_2Au$/Fe Bilayers, Adv. Funct. Mater. **26**, 5884-5892. (2016).

[31] M. Jourdan, H. Bräuning, A. Sapozhnik, H. J. Elmers, H. Zabel, and M. Kläui, Epitaxial $Mn_2Au$ thin films for antiferromagnetic spintronics, J. Phys. D: Appl. Phys. **48**, 385001 (2015).

[32] See the Supplemental Material for experimental details, magnetic properties of $Mn_2Au$ (101) and (204) films, different thickness (103) orientated $Mn_2Au$ switching behavior, raw data for XAS, pole figure of $Mn_2Au$ (103), (101) and (204) films, Joule heat and Oster field simulation and magnetocrystalline anisotropy energy parameters.

[33] R. Masrourab, E. K. Hlilc, M. Hamedound, A. Benyoussefbde, A. Boutaharf, and H. Lassrif, Antiferromagnetic spintronics of $Mn_2Au$: An experiment, first principle, mean field and series expansions calculations study, J. Magn. Magn. Mater. **393**, 600-603 (2015).

[34] O. J. Lee, L. Q. Liu, C. F. Pai, Y. Li, H. W. Tseng, P. G. Gowtham, J. P. Park, D. C. Ralph, and R. A. Buhrman, Central role of domain wall depinning for perpendicular

magnetization switching driven by spin torque from the spin Hall effect, Phys. Rev. B **89**, 024418 (2013).

[35] B. Cui, F. Li., C. Song, J. J. Peng, M. S. Saleem, Y. D. Gu, S. N. Li, K. L. Wang, and F. Pan. Insight into the antiferromagnetic structure manipulated by electronic reconstruction, Phys. Rev. B **94,** 134403 (2016).

[36] L. J. Zhang, Z. J. Xu, X. Z. Zhang, H. N. Yu, Y. Zou, Z. Guo, X. J. Zhen, J. F. Cao, X. Y. Meng, J. Q. Li, Z. H. Chen, Y. Wang, and R. Z. Tai, Latest Advances in Soft X-ray Spectromicroscopy at SSRF, Nucl. Sci. Tech. **26**, 040101 (2015).

[37] A. A. Sapozhnik, R. Abrudan, Yu. Skourski, M. Jourdan, H. Zabel, M. Kläui, and H. J. Elmers, Manipulation of antiferromagnetic domain distribution in $Mn_2Au$ by ultrahigh magnetic fields and by strain, Phys. Status Solidi RRL. **11**, 1–4 (2017).

[38] S. Yu. Bodnar, L. Šmejkal, I. Turek, T. Jungwirth, O. Gomonay, J. Sinova, A. A. Sapozhnik, H. J. Elmers, M. Kläui, and M. Jourdan, Writing and Reading antiferromagnetic $Mn_2Au$: Néel spin-orbit torques and large anisotropic magnetoresistance, Nat. Commun. **9**, 348 (2018).

[39] M. Meinert, D. Graulich, and T. Matalla-Wagner, Key role of thermal activation in the electrical switching of antiferromagnetic $Mn_2Au$, arXiv:1706.06983.

**Figure Captions**

FIG. 1. Crystalline characterization of quasi-epitaxial $Mn_2Au$ films. XRD spectra (left panel) and concomitant HRTEM images (right panel) for $Mn_2Au$ (a) (103), (b) (101), (c) (204) films deposited on MgO (111), $SrTiO_3$ (100), and MgO (110) substrates, respectively.

FIG. 2. Magnetic properties of $Mn_2Au$ (103) films. Magnetic field was applied in plane, i.e. parallel to MgO substrate (111) plane, [110] edge. (a) Magnetization curve of 40 nm $Mn_2Au$ (103) films grown on MgO (111) substrates at 300 K. (b) Magnetic hysteresis loop of $Mn_2Au$ (10 nm) / permalloy (5 nm) bilayer at 20 K. (c) Schematic of 3 × 3 unit cells of $Mn_2Au$, where the $Mn_2Au$ (103) plane is highlighted. Magnetic moments of neighboring sublattices are opposite.

FIG. 3. Schematic of the switching measurement and current-driven switching of $Mn_2Au$ (103) films. (a) Schematic of the star device and switching measurement geometry. Writing current pulses are applied along red ($J_{write}1$) and blue arrows ($J_{write}2$) alternatively, corresponding to two orthogonal directions, [010] and [$\bar{3}01$], for $Mn_2Au$ (103) films. Reading current is marked as the green arrow, while the concomitant Hall resistance is detected in its transverse direction. (b) Sketch of alternative writing current pulses along two orthogonal directions, $J_{write}1$ (red) and $J_{write}2$ (blue) of ~$10^7$ A cm$^{-2}$, and the reading current of ~$10^5$ A cm$^{-2}$ (green) after each writing current pulse. (c) Hall resistance ($R_{Hall}$) change as a function of the number of writing current pulses. For $Mn_2Au$ (103), five successive writing current pulses are applied along [010] and [$\bar{3}01$] (blue) axes alternatively. After each writing current pulse, Hall resistance is recorded during applying a reading current. The variation of $R_{Hall}$ are shown by red and blue squares for the writing current pulse along [010] and [$\bar{3}01$], respectively. (d) Comparison of the variation of $R_{Hall}$ with different magnitude of writing current pulses. (e)

Comparison of the variation of $R_\text{Hall}$ with variable writing pulses´ width. (f) The measurements of $R_\text{Hall}$ are identical as (c), but both of the writing current and reading current directions are rotated 45° from their counterparts in (c). All of the measurements are done at sample temperature of 300 K and a constant background of Hall resistance is subtracted.

FIG. 4. Schematic of XMLD measurement and XMLD signals at different in-plane rotation angles ($\theta$). The left panel shows the XMLD signals at four typical angles between horizontal polarized x-ray direction and [010] axis of $Mn_2Au$ ($\theta$ = 0°, 30°, 45°, and 60°). The right panel displays the sketch for the rotation of the horizontal ($/\!/$, grey solid arrows) polarized x-ray, and vertical ($\perp$, blue solid arrows) polarized x-ray, while the $Mn_2Au$ [010] axis is set at $\theta$ = 0°. The XMLD signals are measured at 300 K.

FIG. 5. SOT switching and spin texture measurements of $Mn_2Au$ (101) films. (a) Dependence of $R_\text{Hall}$ variation on the number of writing current pulses. Three successive current pulses of $J_\text{write}$ = 2.00 × 10$^7$ A cm$^{-2}$ are applied along [13$\bar{1}$] and [$\bar{1}$31] axes alternatively. After each writing current pulse, $R_\text{Hall}$ is recorded during applying the reading current of $J_\text{read}$ = 4.17 × 10$^5$ A cm$^{-2}$, which is denoted by red and blue squares corresponding to the writing current along [13$\bar{1}$] and [$\bar{1}$31] axes, respectively. (b) The measurement procedure is same as (a), but the measurement configuration (both the writing current pulse and reading current) is rotated 45° from (a). (c) Schematic of XMLD measurement and XMLD signals under different in-plane rotation angles ($\theta$). The left panel shows the XMLD signals at three typical angles between horizontal polarized x-ray direction and [13$\bar{1}$] axis of $Mn_2Au$ ($\theta$ = 0°, 30°, and 45°). The right panel displays the sketch for the rotation of the horizontal ($/\!/$, grey solid arrows) polarized x-ray, and vertical ($\perp$, blue solid arrows) polarized x-ray, while the $Mn_2Au$ [13$\bar{1}$] axis is set at $\theta$ = 0°.

FIG. 6. SOT switching and spin texture measurements of $Mn_2Au$ (204) films. (a) Dependence of $R_{Hall}$ on the number of writing current pulses. Four successive current pulses of $J_{write} = 2.00 \times 10^7$ A cm$^{-2}$ are applied along [010] and [$\bar{2}$01] axes alternatively. After each writing current pulse, $R_{Hall}$ is recorded during applying the reading current of $J_{read} = 4.17 \times 10^5$ A cm$^{-2}$, which is denoted by red and blue squares corresponding to the writing current along [010] and [$\bar{2}$01] axes, respectively. (b) The measurement procedure is same as (a), but the measurement configuration (both the writing current pulse and reading current) is rotated 45° from (a). (c) Schematic of XMLD measurement and XMLD signals under different in-plane rotation angles ($\theta$). The left panel shows the XMLD signals at three typical angles between horizontal polarization direction and [010] axis of $Mn_2Au$ ($\theta$ = 0°, 30°, and 45°). The right panel displays the sketch for the rotation of the horizontal ($/\!/$, grey solid arrows) polarized x-ray, and vertical ($\perp$, blue solid arrows) polarized x-ray, while the $Mn_2Au$ [010] axis is set at $\theta$ = 0°.

FIG. 7. MAE as a function of magnetization direction ($\omega$, $\varphi$) in $Mn_2Au$ (103), (101) and (204) orientation films. (a) MAE of all space direction in $Mn_2Au$ (103), (101) and (204) orientation films. (b) MAE distribution as a function of in-plane angle $\theta$ with magnetization lying in (103), (101) and (204) planes, respectively. (c) Schematic of relationship between easy axes and current directions. Hollow double arrows represent the antiferromagnetic easy axes and solid thick arrows represent the current directions. The gold plane represents (001) plane and green arrow represents the projection of [100] axis in (103), (101) and (204) planes, respectively.

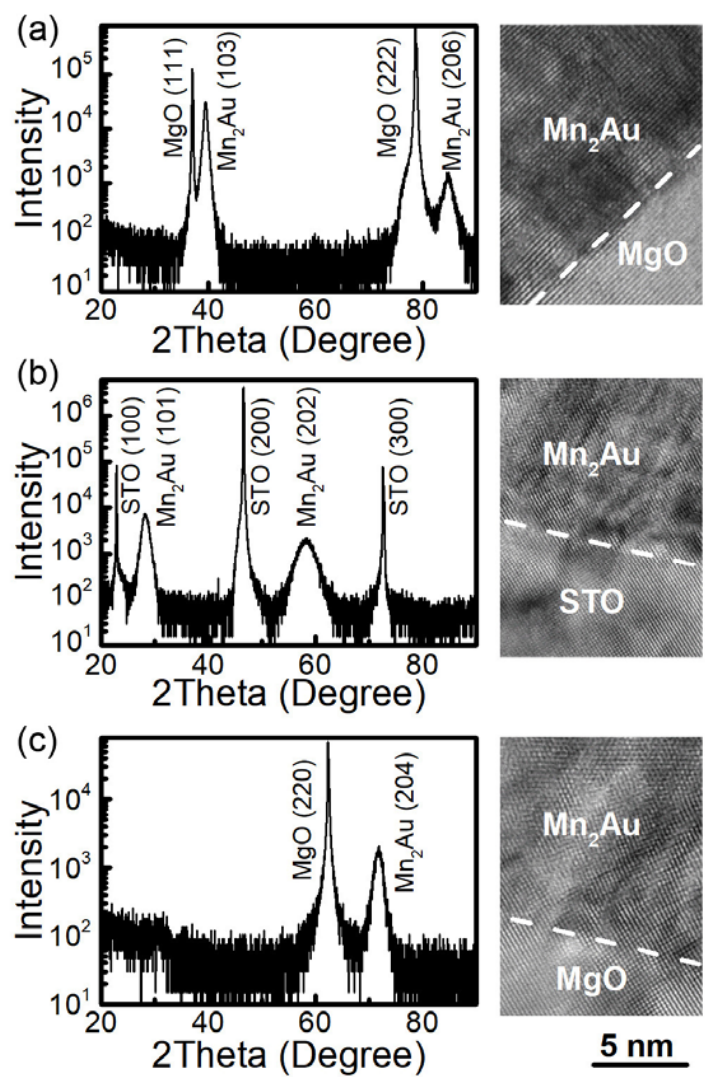

Zhou *et al.* FIG. 1.

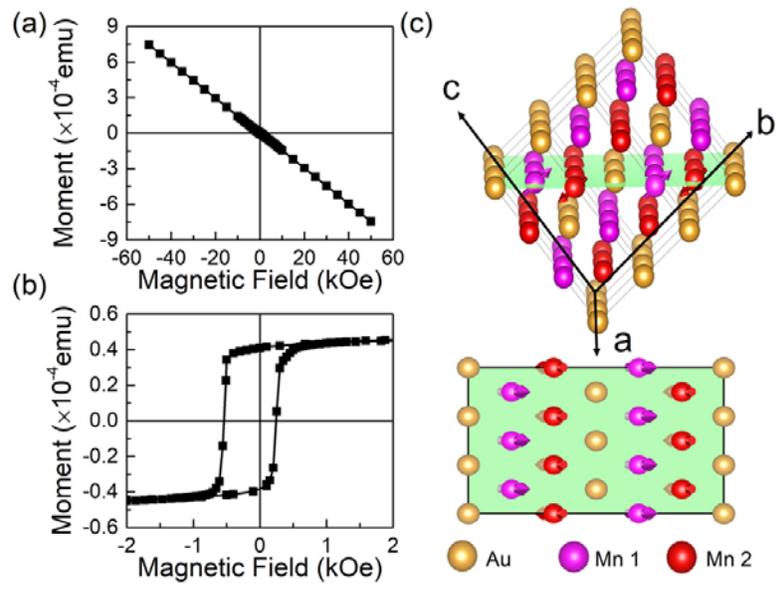

Zhou *et al.* FIG. 2.

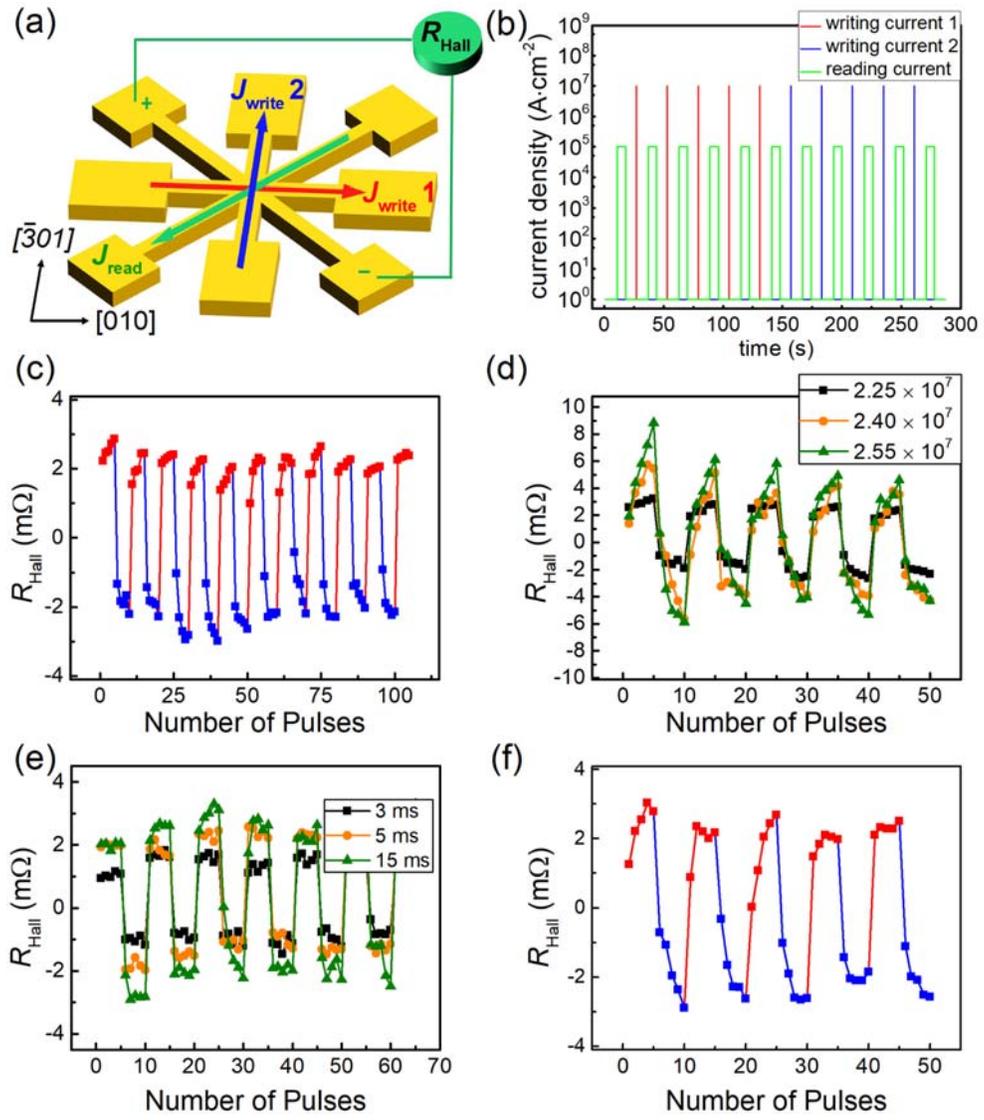

Zhou *et al.* FIG. 3.

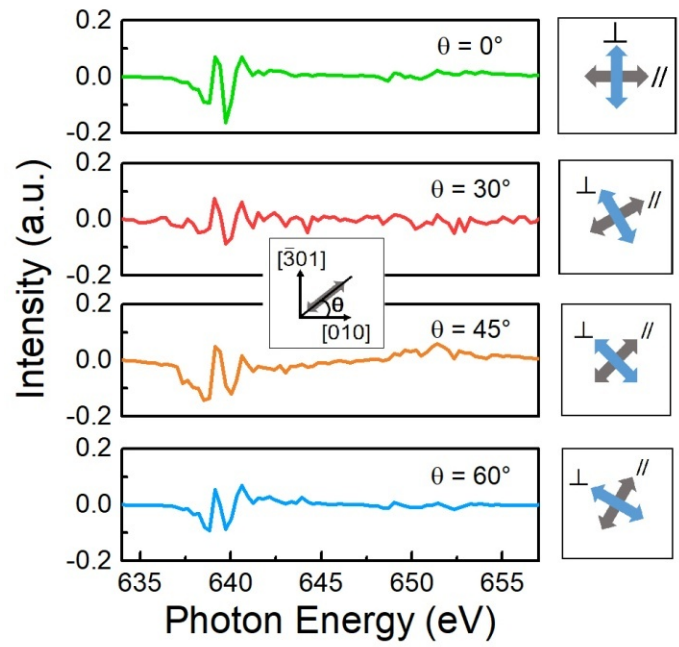

Zhou *et al.* FIG. 4.

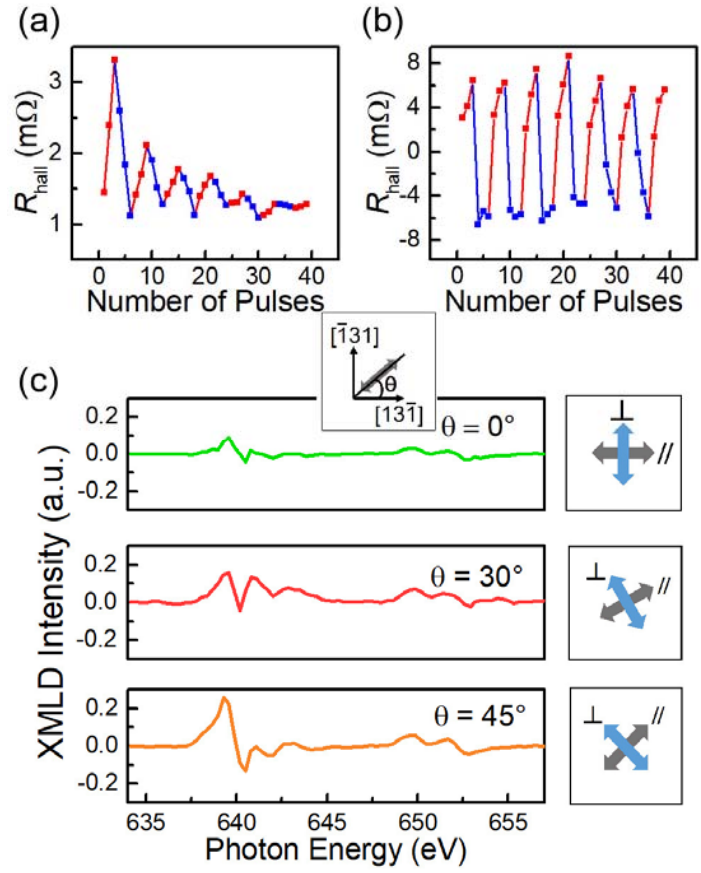

Zhou *et al.* FIG. 5.

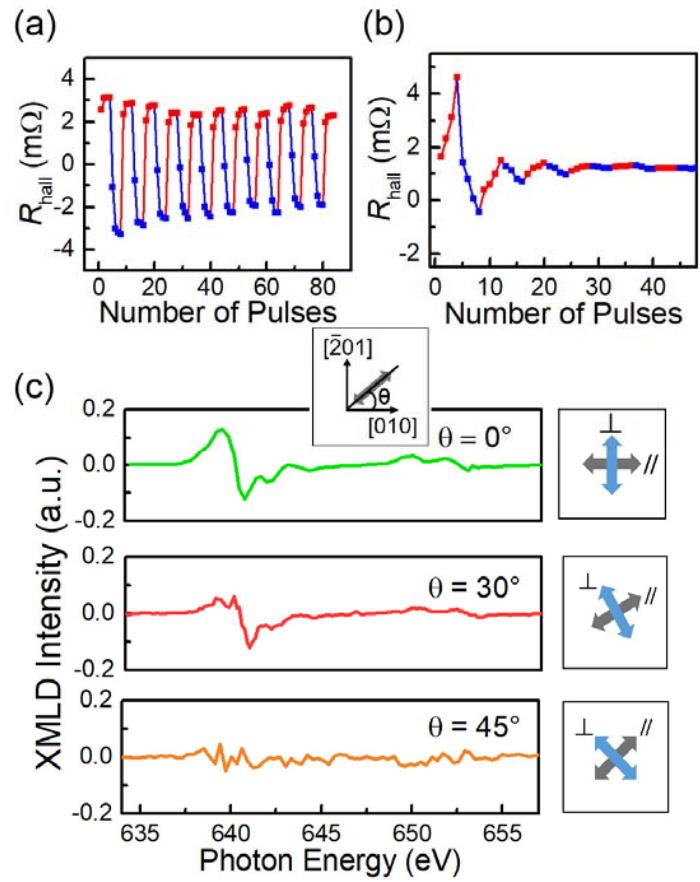

Zhou *et al.* FIG. 6.

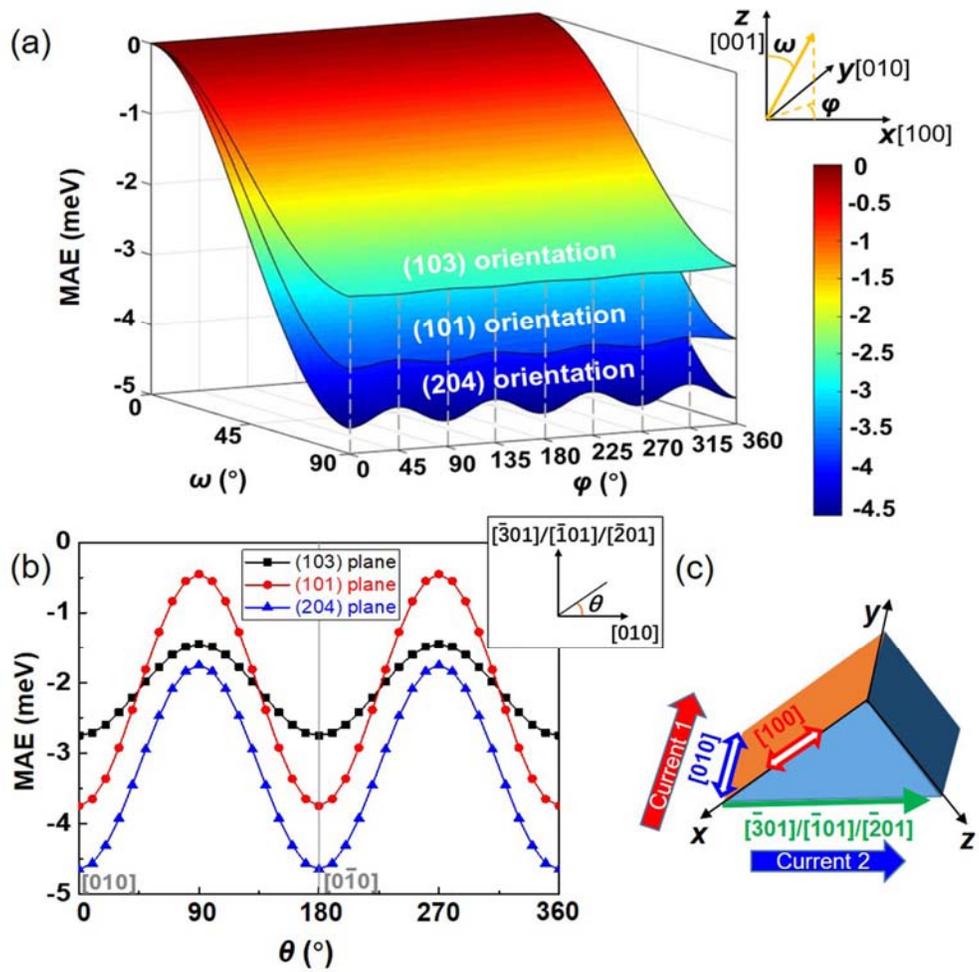

Zhou *et al.* FIG. 7.